\documentclass{kluwer}    

\newdisplay{guess}{Conjecture}

\begin{document}                                                                                   
\begin{article}
\begin{opening}         
\title{Collisions and Mergers of Disk Galaxies: Hydrodynamics of Star 
Forming Gas} 
\author{Susan A.\surname{Lamb}\email{slamb@uiuc.edu}\thanks{Department of Physics, Loomis Laboratory, 1110 W. Green Street, Urbana, IL 61801, USA}}
\author{Nathan C.\surname{Hearn}} 
\runningauthor{S. A. Lamb \& N. C. Hearn}
\runningtitle{Collisions and Mergers of Disk Galaxies}
\institute{University of Illinois}

\begin{abstract}
We summarize the results of numerical simulations of colliding gas-rich disk galaxies in which the impact velocity is set parallel to the spin axes of the two galaxies. The effects of varying the impact speed are studied with particular attention to the resulting gaseous structures and shockwave patterns, and the time needed to produce these structures. The simulations employ an N-body treatment of the stars and dark matter, together with an SPH treatment of the gas, in which all components of the models are gravitationally active. The results indicate that for such impact geometries, collisions can lead to the very rapid formation of a central, rapidly rotating, dense gas disk, and that in all cases extensive star formation is predicted by the very high gas densities and prevalence of shocks, both in the nucleus and out in the galactic disks. As the dense nucleus is forming, gas and stars are dispersed over very large volumes, and only fall back towards the nucleus over long times. In the case of low impact velocities, this takes an order of magnitude more time than that needed for the formation of a dense nucleus. 
\end{abstract}


\end{opening}           

\section{Methods and Models}

In this paper we present a summary of the dominant, large-scale morphologies generated in the gas due to collisions between two co-rotating, gas-rich disk galaxies that have initial velocity vectors parallel to their rotation axes, which are also parallel to each other. We are particularly interested in exploring how quickly a joint nucleus is formed in these 'pancaking' collisions, and what the dynamics of the outer gas might be over time. Here we summarize the results of simulations in which we varied the initial relative velocity for one specific impact parameter of approximately 0.36R, where R is the disk radius, for two equal mass galaxies. These simulations are part of a larger set that will be presented in a subsequent paper, and they complement the studies of colliding disk galaxies of Mihos and Hernquist (1996).

Our simulations employ an N-body/SPH code and initial models as described in Gerber, Lamb \& Balsara (1996). Each galaxy is modeled as an exponential disk of stars, a co-extensive disk of gas, and a halo of mostly dark matter and a few stars, all gravitationally active. These are represented, respectively, by 50,000 star particles, 45,000 SPH gas 'particles,' and 50,000 star/dark matter particles, with the relative masses of 9:1:25. There is no explicit "bulge" component in these galaxy models, so they are best thought of as representing Sc galaxies. The masses and dimensions of the model galaxies can be scaled to those of real galaxies and thence the computational timescale of a simulation and velocities can be related to physical quantities. The physical scaling is discussed in Gerber et al. (1996). The gas is assumed to be isothermal, which is a good approximation under many circumstances, and provides an adaquat description for investigations of flow velocities and the build-up of dense regions in the gas. We track regions of high density and shocks in the gas which we have shown to correspond closely with real star-forming regions (see Lamb, Hearn, \& Gao 1998, and Hearn \& Lamb, 2001). This allows us to compare the results of the simulations with observations of real galaxies, and to explore the likely 3-D structure and kinematics of those systems. 

In these simulations, we varied the initial relative velocity, from about  80 km/s to 300 km/s, scaled to the Milky Way Galaxy's mass and radius. We limit our discussion to the results for one impact parameter, approximately 0.36R, for four different initial relative velocities. In terms of an initial "parabolic" relative velocity of impact, V, for which the two galaxies are marginally bound after impact, our chosen initial velocities are approximately 0.97V, 0.73V, 0.49V, and 0.24V in magnitude. Scaled to the mass and radius of the Milky Way Galaxy, these initial speeds are approximately 312, 234, 156, and 78 km/s, which range across the values of those commonly observed for galaxies in groups and which might be expected to lead to a merger within $10^9$ years of initial impact. We included the very low impact velocity collisions because we wished to investigate collisions which might give rise to a merged galactic nucleus in a relatively short time. The choice of an impact parameter of 0.36R provides a sufficiently off-center collision to produce marked effects of the galactic orbital angular momentum.

\section{Emerging Structure and Velocity Patterns}

The four simulations were run for a computational time equivalent to approximately 0.5 billion years, with the adopted physical scaling. In this time, the galaxies in the fastest collision do not merge. In the three slower cases, the galactic nuclei do merge, in the sense that a central object is formed. However, this 'nucleus' is neither spherical nor dynamically settled. It continues to increase in mass throughout the simulation, as gas and stars flow back towards the joint center from the halo, and, in the cases of the 0.73V and 0.49V collisions, the distribution of material in the nucleus is not stable, but oscillates along three axes. 

The slower the impact speed, the more angular momentum ends up in the central, nuclear gas disk, which forms around the merging nucleus in the three slower impact velocity cases. In the 'slowest' case, the two nuclei barely survive the collision, and a central nucleus forms rapidly from gas and stars falling towards the center. In this case, the central gas disk has considerable angular momentum, drawn from both the original disks and the galactic orbit. Consequently the final disk does not lie parallel to the original planes of the two galaxy disks, but rather its orientation is mostly determined by the orbital angular momentum. The original disk angular momentum contributes 'thickness' to the resulting nuclear gas disk. The central high-density nucleus has a gaseous bar, surrounded by a lower density gaseous ring.

The fastest impact velocity simulation gives rise to two distinct, ring-like galaxies that sit apart at constant separation within the overall joint, elongated halo during our simulation. As in disk galaxies impacted by an elliptical (Lamb et al. 1998), once the ring forms, it expands outward through the disk over time, and is very likely the source of intense star formation. Two real ring galaxies formed from such a collision would have an apparent size somewhat smaller than the extent of the original galaxies, if viewed in the blue light of stars newly formed in the dense rings of gas. At the same time, very extended tails of low-density gas are produced that might be detectable as either extended hot plumes, if the material has passed through shocks in the gas during the collision, or as extended regions of atomic hydrogen, due to gas being flung out of the parts of the disks that did not collide directly, and were thus not shock heated. Here, we do not model the temperature of the gas, so the details of these hot and cold structures must await further simulations with appropriate physics included. Further investigations are also allowing us to study the details of the relative velocity above which merger is delayed for very long periods, that is that impact velocity above which merger does not take place. The results of this study will be presented in a forthcoming paper (Lamb, Hearn, and Morgan, in preparation).

The two intermediate impact velocity simulations produce rather similar resulting morphologies and velocity structures. As in all collisions, the nuclei contract as the galaxies overlap, due to the increased gravitational potential, then they expand as they move apart with the galaxies as a whole. The gas is more affected by the interaction than the stellar component because of its collisional nature. Streamers of gas tend to follow the nuclei as they move apart. These nuclei retain their distinct identities throughout the initial collision, but they rapidly fall back towards each other, producing an oscillating central structure that includes a relatively small, rapidly rotating nuclear gas disk that forms where the two gas streams collide. As the newly forming joint nucleus is fed with inflowing streams of material traveling at several hundred kilometers per second, the outer material is both rotating and expanding. Lobes of gas, expected to be heated up to $10^7$ degrees by the strong shocks experienced in the collision, where relative gas velocities can become very high due to gravitational affects, expand outwards roughly perpendicular to the original disk plane, while un-shocked, cooler gas escapes in the orthogonal directions.

\section{Conclusions}

In the types of collision studied here, a variety of structures can be produced in the gas relatively quickly. In slow encounters (less than approximately 3V/4), a compact central disk and an expanding ring of high density, or a more spherical bulge with a rapidly rotating embedded disk, are likely sites of intense star formation. In higher velocity cases (our 0.97V simulation is an example) the two galaxies stay distinct for much longer periods, each forms a dense gas ring, but has little gas internal to this. In the former cases, the final central disk is not oriented parallel to the original planes of the two galactic disks. Its orientation depends mostly on the orbital angular momentum of the colliding galaxies. Much of the initial galactic material falls back towards the nuclei. Ultimately, much of the outer gas settles into a warped, thick disk, whose outer reaches are still expanding radially outwards at the end of our simulations. Some gas is disbursed widely. In a cluster environment, this gas could be lost to the system and would exist in the overall gravitational potential of the cluster.

\acknowledgements

The authors wish to thank John Morgan, Soojin Kwan, and Chris Lesher for help with running and analyzing the numerical simulations, and they also wish to acknowledge the support of DOE/LLNL contract B506657 and NASA award GO0-1166B.


\end{article}

\end{document}